\documentclass[pra,aps,a4paper,showpacs,superscriptaddress,twocolumn]{revtex4-1}
\usepackage{amsmath}
\usepackage{amsfonts}
\usepackage{amssymb}
\usepackage{graphicx}
\usepackage{dcolumn}
\usepackage{bm}
\begin{document}

\title{Controlling phase separation of a two-component
Bose-Einstein condensate by confinement}
%\author{Lin Wen}
%\author{J. M. Zhang}\email{jmzhang@iphy.ac.cn}
%\author{W. M. Liu}\email{wmliu@iphy.ac.cn}
%\affiliation{Institute of Physics, Chinese Academy of Sciences, Beijing 100080, China}

\author{L. Wen}
\affiliation {Institute of Physics, Chinese Academy of
Sciences, Beijing 100080, China}
\author{W.~M. Liu}
\affiliation {Institute of Physics, Chinese Academy of
Sciences, Beijing 100080, China}
\author{Yongyong Cai}\email{matcaiy@nus.edu.sg}
\affiliation {Department of Mathematics, National University of Singapore,
 Singapore 119076}
\author{J.~M. Zhang}\email{jmzhang@iphy.ac.cn}
\affiliation {Institute of Physics, Chinese Academy of
Sciences, Beijing 100080, China}
\author{Jiangping Hu}\email{hu4@purdue.edu}
\affiliation {Institute of Physics, Chinese Academy of
Sciences, Beijing 100080, China} \affiliation {Department
of Physics, Purdue University, West Lafayette, IN 47906}

\begin{abstract}
We study the effect of kinetic energy on the phase separation and phase transition of a two-component Bose-Einstein condensate in the presence of external confinement. The commonly accepted condition  for the phase separation, $g_{11}g_{22}
<g_{12}^2$ where $g_{11}$, $g_{22}$, and $g_{12}$ are the intra- and inter-
component interaction strengths respectively, is only valid when kinetic energy is negligible and external confinement is nonexistent. Taking a $d$-dimensional infinitely deep square well potential of width $L$ as an example, a simple scaling analysis shows that regardless of the condition $g_{11}
g_{22} <g_{12}^2$, if $d=1$ ($d=3$), phase separation will be suppressed as $L\rightarrow 0$
($L\rightarrow \infty$)  and if $d=2$, the width $L$ is irrelevant but again phase
separation can be partially, or even completely suppressed. Moreover,  the miscibility-immiscibility
transition is turned from a first-order one into a
second-order one when kinetic energy is considered. All these results carry over to
$d$-dimensional harmonic potentials, where the harmonic
oscillator length $\xi_{ho}$ plays the role of $L$. Our
finding provides a scenario of controlling the
miscibility-immiscibility transition of a two-component
condensate by changing the confinement, instead of the
conventional approach of changing the values of the $g$'s.
\end{abstract}

\pacs{03.75.Mn, 05.30.Jp}
\maketitle

\section{Introduction}
Phase separation is a ubiquitous phenomenon in nature
\cite{soft,dagotto}. A most prominent example familiar to
everyone is that oil and water do not mix. Besides that,
the phenomenon of water in coexistence with its vapor can
also be understood as a type of phase separation
\cite{huang}. In general, two phases mix or not depending
on which configuration minimizes the energy or free energy
of the whole system. With the realization of Bose-Einstein
condensation in ultracold atomic gases, another example of
phase separation is offered by two-component Bose-Einstein
condensates (BECs)
\cite{myatt,stamper,hall,modugno,mudrich}. In such a
system, phase mixing or separation means the two
condensates overlap or not spatially, which correspond to
different interaction energies. A widely accepted condition
for phase separation, which is based on the consideration
of minimizing the interaction energy
\cite{pethick,pitaevskii}, is given by
\begin{equation}\label{cond}
g_{12}>\sqrt{g_{11}g_{22}}.
\end{equation}
Here $g_{11}$ and $g_{22}$ are the intra-component
interaction strengths of components $1$ and $2$,
respectively, while $g_{12}$ is the interaction strength
between them \cite{positive}. This condition is intuitively
reasonable since if the inter-component interaction is too
strong, the two components would like to get separated from
each other. Experimentally, controlled
miscibility-immiscibility transition of a two-component BEC
based on the idea of adjusting the values of the $g$'s
using Feshbach resonance and so as to get (\ref{cond})
satisfied or not has been demonstrated recently
\cite{papp08,thalhammer08}.

Now the point is that though the condition above is very
appealing in its simplicity and usefulness, it has great
limitations. In its derivation, the condensates are assumed
to be uniform and the kinetic energy associated with the
boundary/interface layers is neglected. The problem is then
reduced to minimizing the total interaction energy, or more
specifically, to weighing the inter-component interaction
against the intra-component interaction. This approximation
is legitimate if the widths of the boundary/interface
layers are much smaller than the extension of the
condensates, or in other words, if the boundary/interface
layers are well defined. However, this condition is not
necessarily satisfied in all circumstances. Actually, some
simple scaling analysis may tell us when it will fail.
Consider a condensate trapped in a $d$-dimensional
container of size $L$. The characteristic (average) density
of the condensate is on the order of $L^{-d}$. According to
the mean-field (Gross-Pitaevskii) theory, the healing
length of the condensate, which determines the widths of
the boundary/interface layers, will be on the order of
$L^{d/2}$ \cite{pethick,pitaevskii}. Thus we see that in
one and three dimensional cases, it makes sense to say
boundary/interface layers only in the limits of
$L\rightarrow \infty$ and $L\rightarrow 0$, respectively.
In the opposite limits, the ``boundary/interface'' layers
overtake the condensates themselves in size, which signals
that the kinetic energy will dominate the interaction
energy and should no longer be neglected. The two
dimensional case is more subtle in that the widths of the
boundary/interface layers scale in the same way with the
sizes of the condensates, which at least means that the
kinetic energy should not be neglected \textit{a priori}.

The analysis above indicates that the kinetic energy is
likely to play a vital role in determining the
configuration of a two-component BEC. Moreover, we note
that the kinetic energy acts against the inter-component
interaction. The latter is responsible for phase separation
while the former tries to expand the condensates and thus
favors phase mixing. Therefore, it is expected that phase
separation can be suppressed by the kinetic energy in some
circumstances even if the condition (\ref{cond}) is
satisfied \cite{rabi}. Notably, according to the argument
above, the significance of the kinetic energy can be
controlled by changing the size of the container. That is,
the phase mixing-demixing transition can be controlled by a
geometrical method, instead of the mechanical method of
changing the values of the $g$'s, which is based on
(\ref{cond}) and is demonstrated in
Refs.~\cite{papp08,thalhammer08}.

\section{A two-component BEC in an infinitely deep square well potential}

The considerations above have led us to investigate the
scenario of suppressing phase separation in a two-component
BEC by kinetic energy. We will start from the simplest and
most generic case of a two-component BEC in a
$d$-dimensional infinitely deep square well potential (of
width $L$).  The Dirichlet boundary condition implies that
the condensate wave functions must be non-uniform and the
kinetic energy is at least on the order of $L^{-2}$. On the
contrary, inside the well, the potential energy is zero.
Therefore, we have a pure competition between the kinetic
energy and the inter-component interaction energy, if the
intra-component interactions are set zero [note that in
this case, condition (\ref{cond}) is satisfied]. In this
simplest model, in all dimensions ($d=1$, 2, 3), we do
observe that phase separation can be completely suppressed
by the kinetic energy in some regime. Of course, different
dimensions have different features. But all these effects
and features carry over to the more realistic case of
$d$-dimensional harmonic potentials.

In the mean-field theory and at zero-temperature, the
energy functional of a two-component BEC in a
$d$-dimensional infinitely deep square well potential
$\Omega= [-L/2,+L/2]^d$ is of the form
\begin{eqnarray}\label{fun1}
  E[\psi_1,\psi_2] &=& \int_{\Omega} d \vec{r} \bigg\{
 \sum_{\alpha=1,2} \frac{N_\alpha\hbar^2}{2m_\alpha}|\nabla \psi_\alpha|^2 \quad \nonumber \\
  && \quad + \frac{1}{2} \sum_{\alpha,\beta=1,2} g_{\alpha\beta}
  N_\alpha N_\beta |\psi_\alpha|^2|\psi_\beta|^2
  \bigg\}.
\end{eqnarray}
Here the two condensate wave functions are normalized to
unity $\int_\Omega d \vec{r} |\psi_{1,2}|^2=1$, and
$\psi_{1,2}=0$ on the boundary. Note that throughout this
paper we are only concerned with the ground configuration
of the system, therefore all the wave functions can be
taken to be real and positive. The parameters $g_{11}$,
$g_{22}$, and $g_{12}=g_{21}$ are the effective intra- and
inter-component interaction strengths. Finally, $N_{1,2}$
and $m_{1,2}$ are the atom numbers and atom masses of the
two species, respectively. Now we should note that for an
arbitrary set of parameters, in the ground configuration,
almost definitely, the two wave functions do overlap but do
not coincide with each other (this can be easily understood
in terms of the Gross-Pitaevskii equations for
$\psi_{1,2}$). In this case, it is far from trivial to
distinguish phase separation and phase mixing. A method
proposed in \cite{malomed} is to consider the centers of
mass of the two condensates:
\begin{equation}\label{centerofmass}
    \vec{r}_{m\alpha }=\int_\Omega d \vec{r}
    |\psi_\alpha|^2 \vec{r}, \quad \alpha=1,2.
\end{equation}
This idea is motivated by the observation that in some
regime, both the two condensates are symmetric with respect
to the origin while in other regime, both of them are
asymmetric with respect to the origin, and more
importantly, they are shifted in opposite directions
\cite{malomed}. Apparently, the former case is with
$\vec{r}_{m1 }=\vec{r}_{m2 }=0$ and it is appropriate to
call it phase-mixed while the latter case is with
$\vec{r}_{m1 }\neq 0 \neq\vec{r}_{m2 }$ and it is
appropriate to call it phase-separated. Therefore, the
offset between the two centers of mass $\vec{r}_{m1
}-\vec{r}_{m2 }$ can serve as an order parameter for the
miscibility-immiscibility transition of the system.

Though this order parameter works well for a general case,
we will not use it much in this paper. Actually, instead of
studying a general case, we shall focus on the symmetric
energy functional case, i.e., the case when $m_1=m_2=m$,
$N_1=N_2=N$, and $g_{11} =g_{22}$. The reason is that this
special case not only captures all the essential physics,
but also has an extra merit. That is, now it is possible to
have $\psi_1=\psi_2$, which corresponds to a completely
mixed configuration. Therefore, in this special case, an
appropriate order parameter is the overlap between the two
condensate wave functions (or more precisely, $1-\eta$, if
phase separation is concerned):
\begin{equation}\label{overlap}
\eta=\int_\Omega d \vec{r} \psi_1 \psi_2,
\end{equation}
which takes values between 0 and 1. If $\eta\ll 1$, it
would be fair to say the system shows phase separation.
Otherwise, if $\eta$ is close to 1, or more precisely if
$1-\eta \ll 1$, it would be fair to say the system shows
phase mixing. In the intermediate case, the system is
partially phase-separated and partially phase-mixed.

Now make the transform $\psi_{1,2}(\vec{r})= L^{-d/2}
\phi_{1,2}(\vec{x})$ with $\vec{r}= L \vec{x}$. Then
$\int_{\Omega_0} d \vec{x} |\phi_{1,2}|^2=1$ and
$\phi_{1,2}=0$ on the boundary of $\Omega_0$, where
$\Omega_0=[-1/2,+1/2]^d$. In terms of the rescaled wave
functions $\phi_{1,2}$, $\eta=\int_{\Omega_0} d \vec{x}
\phi_1 \phi_2 $, and the energy functional (\ref{fun1}),
under the assumption above, can be rewritten as
\begin{eqnarray}\label{fun2}
  E[\phi_1,\phi_2]=\frac{N\hbar^2}{m L^2} \int_{\Omega_0} d
\vec{x} \bigg\{ \frac{1}{2}|\nabla \phi_1|^2+ \frac{1}{2}|\nabla \phi_2|^2 \ \quad\quad\quad \nonumber \\
   + \frac{1}{2}\left(\beta_{11} |\phi_1|^4 +
\beta_{22} |\phi_2|^4  +2\beta_{12} |\phi_1|^2|\phi_2|^2
\right)
  \bigg\},
\end{eqnarray}
with the reduced dimensionless parameters $\beta_{ij}$
defined as
\begin{equation}\label{beta}
    \beta_{ij} = \frac{Nmg_{ij}}{ \hbar^2 L^{d-2}},\quad
    i,j=1,2.
\end{equation}
These parameters are measures of the importance of the
interactions. In the curl bracket, the coefficients of the
kinetic terms are constant, yet the coefficients of the
interaction terms (the $\beta$'s) scale with $L$ as
$L^{2-d}$. This fact has some important consequences. If
$d=1$, there are two different limits. In the limit of
$L\rightarrow \infty$ (loose confinement), the kinetic
terms are dominated by the interaction terms and thus the
ground state can be determined by simply minimizing the
interaction energy. In this limit, the textbook analysis is
valid and we have phase separation if condition
(\ref{cond}) is satisfied or phase mixing otherwise. In the
opposite limit of $L\rightarrow 0$ (tight confinement), the
kinetic terms will dominate and the two rescaled wave
functions can be well approximated by the ground state of
the square well potential, i.e.,
$\phi_{1,2}(x)\simeq\sqrt{2}\cos (\pi x)$. In this limit,
phase separation will be suppressed whatever the values of
the $g$'s are, even if (\ref{cond}) is fulfilled. The three
dimensional case is the inverse of the one dimensional
case. In the limit of $L\rightarrow 0$, the kinetic terms
are negligible and the criterion of phase separation
(\ref{cond}) is valid. In the other limit of $L\rightarrow
\infty$, the kinetic terms dominate and phase separation is
suppressed regardless of the condition (\ref{cond}). The
two dimensional case is another story. The parameter $L$
simply drops out in the curl bracket. It is no use to
adjust the width of the well to enhance the importance of
the kinetic energy or the interaction energy relatively.
The kinetic and interaction energies should be treated on
an equal footing, which means the analysis leading to
criterion (\ref{cond}) may be invalid.
\begin{figure}[tb]
\includegraphics[ width= 0.48\textwidth]{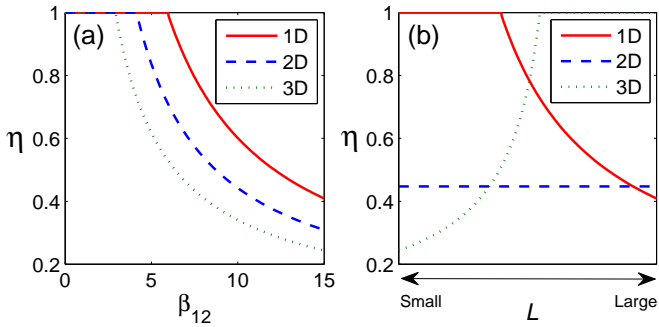}
\caption{(Color online) (a) The overlap factor $\eta $ as a
function of the reduced parameter $\beta_{12}$ [see
Eq.~(\ref{beta})] in different dimensions (infinitely deep
square well potential case, $g_{11}=g_{22}=0$). Note that
for all values of $d$, there exists a critical value
$\beta_{12}^c\neq 0$, below which $\eta$ attains its
maximal possible value $1$. (b) a schematic plot of $\eta$
versus the width of the square well in different
dimensions. Note the counter-intuitive fact that in the
three dimensional case ($d=3$), the stronger we squeeze the
system (the smaller $L$ is), the stronger phase separation
is (the smaller $\eta$ is).\label{fig1}}
\end{figure}

We have checked all these predictions numerically. Note
that on the problem of phase separation, the
intra-component interactions are on the same side as the
kinetic energy---they both try to delocalize the
condensates. Therefore, to highlight the effect of kinetic
energy, we shall set $g_{11}= g_{22}=0$
($\beta_{11}=\beta_{22}=0$), so that the kinetic energy is
the only element acting against phase separation. As we
shall see below, this special case also admits a simple
analytical analysis.

We have solved the ground state of the system in all
dimensions for a given value of $\beta_{12}$ \cite{bao}.
The overlap factor $\eta$ is plotted versus $\beta_{12}$ in
Fig.~\ref{fig1}a. We observe that in all dimensions, there
exists a critical value of $\beta_{12}$ (denoted as
$\beta_{12}^c$), below which the two condensates wave
functions are equal ($\eta=1 $). That is, for
$\beta_{12}\leq \beta_{12}^c$, phase separation is
completely suppressed. Above the critical value, phase
separation develops ($\eta<1$) as $\beta_{12}$ increases,
but is still greatly suppressed for a wide range of value
of $\beta_{12}$. It should be stressed that though in
Fig.~\ref{fig1}a the curves of $\eta-\beta_{12}$ are
qualitatively similar to each another for all values of $d$
(the plateau of $\eta=1$ is always located in the direction
of $\beta_{12} \rightarrow 0$), the curves of $\eta-L$ will
be quite different. The reason is that $\beta_{12} \propto
L^{2-d}$. Figure \ref{fig1}b is a schematic plot of $\eta$
versus $L$ in all the three cases. It shows that $\eta$ as
a function of $L$ is monotonically decreasing, constant,
and monotonically increasing in one, two, and three
dimensions, respectively. This means that to suppress phase
separation, in one dimension we should tighten the
confinement, in three dimensions we should loosen the
confinement, while in two dimensions it is useless to
change the confinement. Overall, Fig.~\ref{fig1} confirms
the initial conjecture that kinetic energy can suppress
phase separation.

As a hindsight, we can actually understand why phase
separation can be suppressed in the limits of $L
\rightarrow 0$ in one dimension and $L\rightarrow \infty$
in three dimensions. Consider two different configurations.
The first one is a phase-separated one---the two
condensates occupy the left and right halves of the
container separately. The second one is a phase-mixed
one---the two condensates both occupy the whole space
available and thus overlap significantly. Compared with the
first configuration, the second one costs more
inter-component interaction energy which is on the order of
$L^{-d}$, but saves more kinetic energy which is on the
order of $L^{-2}$. The second configuration (phase-mixed)
is more economical in energy in the limit of $L\rightarrow
0$ and $L\rightarrow \infty$, in the cases of $d=1$ and
$d=3$, respectively. The case of $d=2$ is more subtle and
which configuration wins depends on parameters other than
$L$.

A remarkable fact revealed in Fig.~\ref{fig1} but not so
obvious in our arguments is that in the symmetric case with
$\beta_{11}=\beta_{22}=0$, $\eta=1$ for $\beta_{12}\leq
\beta_{12}^c$, which is on the order of unity. This is a
stronger fact than $\eta \rightarrow 1$ as $\beta_{12}
\rightarrow 0$ as we argued. Actually, the general
observation is that for $\beta_{11}=\beta_{22}>0$, $\eta=1$
for $\beta_{12}$ smaller than its critical value $
\beta_{12}^c$, which is larger than $\beta_{11}$. This fact
has rich meanings. On the one hand, it demonstrates that
the kinetic energy is very effective---phase separation can
be \textit{completely suppressed} by it even if $\beta_{12} >
\beta_{11}=\beta_{22}$, i.e., when (\ref{cond}) is
satisfied. On the other hand, it strongly indicates that as
$\beta_{12}$ crosses the critical value, the system
undergoes a second order phase transition which can fit in
the Landau formalism. The picture is that the exchange
symmetry $\phi_1 \leftrightarrow \phi_2 $ of the energy
functional (\ref{fun2}) is preserved for $\beta_{12} <
\beta_{12}^c$, but is spontaneously broken as $\beta_{12}$
surpasses $\beta_{12}^c$.

We have been able to prove the first point rigorously on the mathematical level (see Appendix \ref{appRigor}). However, it is also desirable to develop a physical understanding of the two points. This can be achieved by  
studying a two-component BEC in a double-well
potential (see Appendix \ref{appSpon}) or using a variational
approach \cite{pst6}. We note that in the limit of
$\beta_{12} \rightarrow 0$, $\phi_{1,2}$ both converge to
the (non-degenerate) ground state of a single particle in
the $[-1/2,+1/2]^d$ infinitely deep square well. As
$\beta_{12}$ is turned on, the two wave functions are
deformed and excited states mix in. Because the energies of
the excited states grow up quadratically, we cutoff at the
first excited level and take the following ansatz for the
two condensate wave functions
\begin{equation}\label{trial}
    \phi_1= c_0 \varphi_0 + c_1 \varphi_1, \quad \phi_2= c_0
\varphi_0 - c_1 \varphi_1.
\end{equation}
Here $\varphi_0 $ is the ground state, while $\varphi_1$ is
one of the possibly degenerate first excited states. The
coefficients $c_{0,1}$ are real and satisfy the
normalization condition $c_0^2 + c_1^2=1$. Obviously,
complete phase mixing would correspond to $c_1=0$ while
partial phase separation to $c_1\neq 0$. Our numerical
simulations indicate that (this is also supported by the variational approach itself, see Appendix \ref{appc}) in the two dimensional
case, when phase separation occurs, the two condensates are
shifted either along $x$ or $y$ direction; in the three
dimensional case, when phase separation occurs, the two
condensates are shifted either along $x$ or $y$ or $z$ direction.
This fact motivates us to choose $\varphi_1$ in the
following form
\begin{subequations}\label{trial2}
\begin{eqnarray}
&&d=1: \varphi_1= w_1(x); \\
&&d=2: \varphi_1= w_0(x) w_1(y) \ \text{  or  } \ w_1(x) w_0(y) ;   \\
&&d=3: \varphi_1= w_0(x) w_0(y)w_1(z) \ \text{  or  } \ w_0(x) w_1(y)w_0(z)\quad   \nonumber \\
&&\quad\quad \quad \quad \quad \quad   \text{       or  } w_1(x) w_0(y)w_0(z),\quad\quad
\end{eqnarray}
\end{subequations}
where $w_0(x)=\sqrt{2} \cos(\pi x)$ and
$w_1(x)=\sqrt{2}\sin(2 \pi x)$ are the ground and first
excited states of a single particle in the one dimensional
$[-1/2,+1/2]$ infinitely deep square well potential.
Substituting Eqs.~(\ref{trial}) and (\ref{trial2}) into
(\ref{fun2}), we get the reduced energy functional
$\widetilde{E}=E/(N\hbar^2/mL^2)$ as
\begin{subequations}\label{reduced e}
\begin{eqnarray}
&&d=1: \widetilde{E}(c_1) = (3\pi^2 -5 \beta_{12})c_1^2 + 5 \beta_{12}c_1^4 +\text{const}; \nonumber \\
&&d=2:  \widetilde{E}(c_1)  = \left(3\pi^2 -\frac{15}{2} \beta_{12} \right)c_1^2 + \frac{15}{2} \beta_{12}c_1^4 +\text{const}; \nonumber \\
&&d=3:\widetilde{E} (c_1) = \left(3\pi^2 -\frac{45}{4}
\beta_{12} \right)c_1^2 + \frac{45}{4}\beta_{12}c_1^4 +\text{const}.
\nonumber
\end{eqnarray}
\end{subequations}
These are nothing but the Landau's expression of the free
energy in a second-order phase transition, with $c_1$
playing the role of the order parameter here. We
immediately determine the critical values of $\beta_{12}$
by putting the coefficients of $c_1^2$ to zero.
Specifically, $\beta_{12}^c=\frac{3\pi^2}{5}$,
$\frac{2\pi^2}{5}$, and $\frac{4\pi^2}{15}$ for $d=1$,
$d=2$, and $d=3$, respectively. These values agree with
those extracted from Fig.~\ref{fig1} very well. The
relative errors are within $1$\%, $9$\%, and $19$\%,
respectively. The deviation increases with $d$ because in
higher dimensions, the degeneracy of the excited states
increases and the two-mode approximation in (\ref{trial})
becomes less accurate. In the expressions of
$\widetilde{E}$, we can actually see how the kinetic energy
suppresses phase separation. The term $3\pi^2 c_1^2$ comes
from the kinetic energy difference of the two modes
$\varphi_{1,2}$. Without this term, the critical value
$\beta_{12}^c$ would be zero instead of being finite.

For a general case without the exchange symmetry $\phi_1
\leftrightarrow \phi_2$, the appropriate order parameter is
no longer $\eta$ but $\vec{r}_{m1 }-\vec{r}_{m2 }$.
However, the second order transition picture still holds.
Specifically, $\vec{r}_{m1 }=0=\vec{r}_{m2 }$ for
$\beta_{12}$ smaller than some critical value $
\beta_{12}^c$ which is larger than $\sqrt{\beta_{11}
\beta_{22}}$. Overall, this asymmetric case is more involved than the symmetric case above because there are more parameters. Hopefully, a systematic study will be presented in a follow-up work.

\section{A two-component BEC in a harmonic potential}

So far, we have focused on the ideal case of infinitely
deep square wells. Experimentally, it is harmonic
potentials that are most readily realized. Therefore, it is
necessary to see whether analogous results hold for
harmonic potentials. One concern is that the extra
potential energy may blur the picture. However, after some
similar rescaling, we shall see that all the results
persist.

The energy functional of a two-component BEC in a
$d$-dimensional isotropic harmonic potential is
\begin{eqnarray}\label{fun3}
  \frac{E}{N}=  \int_{R^d} d \vec{r} \bigg\{ \frac{\hbar^2}{2m}\sum_{\alpha=1,2}|\nabla \psi_\alpha|^2+\frac{1}{2}m\omega_d^2 |\vec{r}|^2 \sum_{\alpha=1,2} |\psi_\alpha|^2 \nonumber \\
    + \frac{N}{2}\left(g_{11} |\psi_1|^4 + g_{22} |\psi_2|^4  +2g_{12} |\psi_1|^2|\psi_2|^2 \right)
  \bigg\}.\ \
\end{eqnarray}
Here again we have assumed equal mass and equal number for
the two species. The two condensate wave functions are
normalized to unity, i.e., $ \int d \vec{r}
|\psi_{1,2}|^2=1 $. Now make the transform
$\psi_{1,2}(\vec{r})= \xi_{ho}^{-d/2} \phi_{1,2}(\vec{x})$
with $\vec{r}= \xi_{ho} \vec{x}$, where
$\xi_{ho}=\sqrt{\hbar/m\omega_d}$ is the characteristic
length of the harmonic potential. We have then $\int d
\vec{x} |\phi_{1,2}|^2=1$. In terms of $\phi_{1,2}$, the
energy functional can be rewritten as
\begin{eqnarray}\label{fun4}
  \frac{E}{N \hbar \omega_d}=  \int_{R^d} d \vec{x} \bigg\{ \frac{1}{2}\sum_{\alpha=1,2}|\nabla \phi_\alpha|^2+\frac{1}{2}|\vec{x}|^2 \sum_{\alpha=1,2}|\phi_\alpha|^2\quad \nonumber \\
    + \frac{1}{2}\left(\beta_{11} |\phi_1|^4 + \beta_{22} |\phi_2|^4  +2\beta_{12} |\phi_1|^2|\phi_2|^2 \right)
  \bigg\}.\quad
\end{eqnarray}
Here the reduced interaction strengths are defined as
\begin{equation}\label{beta2}
    \beta_{ij}= \frac{N m g_{ij} \xi^{2-d}_{ho}}{\hbar^2}
\propto \omega_d^{(d-2)/2},\  i,j=1,2.
\end{equation}

We now have a similar situation as before. The importance
of the interactions can be changed by changing the value of
$\xi_{ho}$, which plays the role of $L$ in our previous
example. The interactions will be negligible if $d=1$ and
$\xi_{ho} \rightarrow 0$ or if $d=3$ and $\xi_{ho}
\rightarrow \infty$. In this case, the rescaled wave
functions $\phi_{1,2}$ will be close to the ground state of
the harmonic oscillator, i.e., $\phi_{1,2} \simeq
\pi^{-d/2} \exp(-\vec{x}^2/2)$, and phase separation is
suppressed regardless of the values of the $g$'s. The
interactions will become significant if $d=1$ and
$\xi_{ho}\rightarrow \infty$ or $d=3$ and $\xi_{ho}
\rightarrow 0$. In this case, the kinetic energy can be
neglected and we enter the Thomas-Fermi regime. In this
regime, the criterion (\ref{cond}) will be a faithful one
for phase separation.

\begin{figure}[tb]
\includegraphics[ width= 0.48\textwidth]{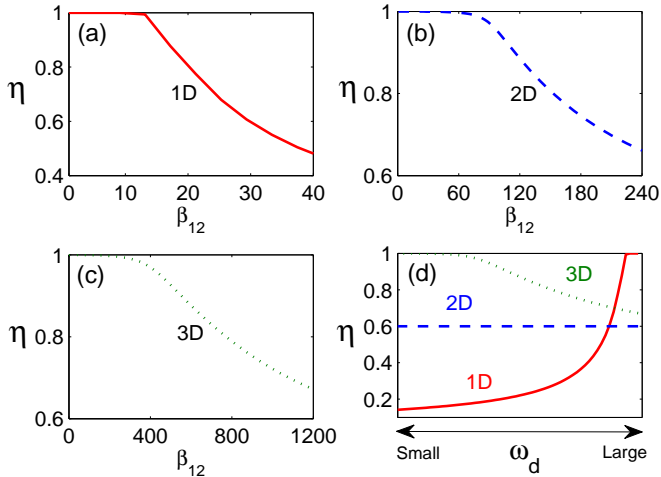}
\caption{(Color online) (a)-(c) The overlap factor $\eta $
as a function of the reduced parameter $\beta_{12}$ [see
Eq.~(\ref{beta2})] in different dimensions (isotropic
harmonic potential case, $g_{11}=g_{22}=g_{12}/1.05$). Note
that for all value of $d$, there exists a critical value
$\beta_{12}^c\neq 0$, below which $\eta$ attains its
maximal possible value $1$. (d) a schematic plot of $\eta$
versus the characteristic frequency $\omega_d$ of the
harmonic potential in different dimensions.\label{fig2}}
\end{figure}

We have verified these predictions numerically. In
Fig.~\ref{fig2}, we have shown the overlap factor
$\eta\equiv \int d \vec{x} \phi_1 \phi_2$ versus the
reduced inter-component interaction strength $\beta_{12}$
in all dimensions (with $g_{11}=g_{22}=g_{12}/1.05$).
Again, we see that phase separation is completely
suppressed for $\beta_{12}$ below some critical value
$\beta_{12}^c$.

Let us now consider the possibility of experimentally
observing the immiscibility-miscibility transition by
adjusting the confinement, e.g., the frequency $\omega_d$.
In cold atom experiments, the harmonic potential is often
of the form $V(\vec{r})= \frac{1}{2}m [\omega_\perp^2 (x^2
+y^2) + \omega_z^2 z^2]$. To get a three dimensional
isotropic potential, we set $\omega_\perp= \omega_z$. An
effectively one (two) dimensional potential can be obtained
in the limit of $\omega_\perp \gg \omega_z$ ($\omega_\perp
\ll \omega_z$). For these three different geometries of the
potential, the interaction strengths (the $g$'s) relate to
the $s$-wave scattering lengths (the $a$'s) as
\begin{subequations}\label{ganda}
\begin{eqnarray}
  g_{ij} \!&=& \frac{4\pi\hbar a_{ij}}{m}, \ \omega_d=\omega_z=\omega_\perp, \ d=3; \\
  g_{ij} \!&=&  \frac{2 \sqrt{2\pi} \hbar^{3/2} \omega_z^{1/2} a_{ij}}{m^{1/2}},\ \omega_d= \omega_\perp\! \ll \omega_z,  \ d=\!2;\quad\\
  g_{ij} \!&=& 2\hbar a_{ij} \omega_\perp, \ \omega_d= \omega_z \ll \omega_\perp, \ d=1.
\end{eqnarray}
\end{subequations}
Using Eqs.~(\ref{beta2}) and (\ref{ganda}), we can study
the possibility of tuning $\beta_{12}$ across the critical
value $\beta_{12}^c$. We study each case individually (the
mass $m$ is taken to be that of $^{23}$Na):

(i) $d=3$. Suppose $N=10^4$, $a_{12}=40 $ $a_{\text{B}}$.
The critical value of $ \omega_d$ is $2\pi\times 560$ Hz,
which can be covered in current experiments.

(ii) $d=2$. Suppose $N=10^{4}$, $a_{12}=40$ $
a_{\text{B}}$, and the transverse frequency $\omega _{\perp
}=2\pi \times 2.6$ Hz. The critical value of the
longitudinal frequency $\omega _{z}$ is $2\pi \times 140$
Hz, which is realizable in current experiments
\cite{papp08}.

(iii) $d=1$. Suppose $N=2\times 10^{3}$, $a_{12}=40$ $
a_{\text{B}}$, and the transverse frequency $\omega _{\perp
}=2\pi \times 130$ Hz. The critical value of the
longitudinal frequency $\omega _{z}$ is $2\pi \times 19$
Hz, which is realizable in current experiments.

Here the number of atoms is one or two orders smaller than
its typical value in experiments. This explains why the
criterion (\ref{cond}) is a reliable one in the experiments
in \cite{papp08,thalhammer08}. They work in a regime where
the kinetic energy is indeed negligible. However, with the
advance of imaging techniques, hopefully future experiments
can work with a relatively small number of atoms and
observe the miscibility-immiscibility transition by
changing the confinement.

\section{Conclusions}

To conclude, we have demonstrated that kinetic energy can
play a vital role in determining the configuration of a
two-component BEC. It renders the empirical condition of
phase separation $g_{11} g_{22} < g_{12}^2$ insufficient
and it also modifies the picture of phase separation. To be
specific, phase separation can be completely suppressed
even if this condition is fulfilled. Moreover, the phase
mixing to phase separation transition is now known to be a
second-order, continuous transition instead of a
first-order, discontinuous one as in the usual view. From
the experimental point of view, our results may provide a
new scenario of controlling the transition of phase
mixing-demixing of a two-component BEC. Instead of
adjusting the interaction strengths, one can just change
the confinement, the characteristic size of the container.

\section{Acknowledgments}
We are grateful to Weizhu Bao, L. You, and C.~H. Lee for helpful
discussions. This work is supported by NSF of China under
Grant No. 11091240226, the Ministry of Science and Technology of China 973 program (2012CV821400) and NSFC-1190024. Y. Cai acknowledges support from the Academic Research Fund of Ministry of
Education of Singapore grant R-146-000-120-112.

\appendix

\section{Rigorous justification}
\label{appRigor}
Here, we consider the energy functional as
\begin{eqnarray}\label{func1}
E[\phi_1,\phi_2]= \int_{\Omega_0} d
\vec{x} \bigg\{ \frac{1}{2}|\nabla \phi_1|^2
+ \frac{1}{2}|\nabla \phi_2|^2\qquad\qquad\qquad \\
   + \frac{1}{2}\left(\beta_{11} |\phi_1|^4 +
\beta_{22} |\phi_2|^4  +2\beta_{12} |\phi_1|^2|\phi_2|^2
\right)\nonumber
  \bigg\},
\end{eqnarray}
where $\Omega_0=[-\frac12,\frac12]^d$ ($d=1,2,3$), $\beta_{11}=\beta_{22}=\beta$. Let $\phi_g$ be the unique positive ground state of the energy functional $E_s[\phi]\equiv E[\phi,\phi]$, and $\mu_g$ be the corresponding chemical potential. The functions $\phi_{1,2}$ are normalized to unity by the usual $L^2$-norm. Let $(\phi^g_1,\phi_2^g)$ be the positive ground state of (\ref{func1}). For $\beta_{12}\leq \beta$, $E[\sqrt{\rho_1},\sqrt{\rho_2}]$ ($\rho_1\equiv |\phi_1|^2$, $\rho_2\equiv |\phi_2|^2$) is strictly convex in $(\rho_1,\rho_2)$ \cite{Lie,2bec}, and the positive ground state is unique, i.e., $\phi_1^g=\phi_2^g=\phi_g$, $\eta=1$. We are going to prove that there exists a critical value $\beta_{12}^c>\beta$ such that when $\beta_{12}<\beta_{12}^c$, there holds $\phi_1^g=\phi_2^g=\phi_g$, i.e., $\eta=1$. From now on, we concentrate on the case of $\beta_{12}\ge\beta$ and assume that $\beta_{12}=\beta+\beta^\prime$, $0\leq\beta^\prime\leq1$. Simple calculation shows that
\begin{align*}
&  E[\phi_1^g,\phi_2^g]-E[\phi_g,\phi_g] = \int_{\Omega_0} d
\vec{x} \sum\limits_{\alpha=1,2}\bigg\{ \frac{1}{2}|\nabla (\phi_\alpha^g-\phi_g)|^2\\
&+(\beta+\beta_{12}) |\phi_g|^2|\phi_\alpha^g-\phi_g|^2
+\frac{\beta-\beta_{12}}{2} \left(|\phi_\alpha^g|^2-|\phi_g|^2\right)^2\\
&+\nabla (\phi_\alpha^g-\phi_g)\cdot\nabla\phi_g
+2(\beta+\beta_{12})|\phi_g|^2\phi_g(\phi_\alpha^g-\phi_g)
\bigg\}
\\&+ \frac{\beta_{12}}{2}\left(|\phi_1^g|^2+|\phi_2^g|^2-2|\phi_g|^2
\right)^2.
\end{align*}
Making use of the Euler-Lagrange equation of $\phi_g$,
\begin{equation}\label{e-l}
\mu_g\phi_g=-\frac12\nabla^2\phi_g+(\beta+\beta_{12})|\phi_g|^2\phi_g,
\end{equation}
denoting $e_\alpha=\phi_\alpha^g-\phi_g$ ($\alpha=1,2$), and noticing
$\int_{\Omega_0}e_\alpha\phi_g=-\frac12\|e_\alpha\|_2^2$, we obtain
\begin{align*}
&  E[\phi_1^g,\phi_2^g]-E[\phi_g,\phi_g] = \int_{\Omega_0} d
\vec{x} \sum\limits_{\alpha=1,2}\bigg\{ \frac{1}{2}|\nabla e_\alpha|^2\\
&+(\beta+\beta_{12}) |\phi_g|^2|e_\alpha|^2
-\frac{\beta^\prime}{2} \left(|\phi_\alpha^g|^2-|\phi_g|^2\right)^2\\
&+2\mu_g \phi_g e_\alpha
\bigg\}+ \frac{\beta_{12}}{2}\left(|\phi_1^g|^2+|\phi_2^g|^2-2|\phi_g|^2
\right)^2.
\end{align*}
Now, the operator $L_g=-\frac{1}{2}\nabla^2+(\beta+\beta_{12})|\phi_g|^2$ admits eigenvalues as $\mu_g<\mu_1\leq\mu_2\leq\cdots$, and the  eigenfunction $\phi_g$ corresponds to $\mu_g$, $w_k\in H_0^1$ with $\|w_k\|_2=1$ corresponds to $\mu_k$ ($k\ge1$). The reason $\phi_g$ is the ground state comes from the positivity of $\phi_g$ and the uniqueness of the positive ground state of $L_g$. Expand $e_\alpha$ as $e_\alpha=c_g^\alpha\phi_g+\sum_{k=1}^\infty c^\alpha_kw_k$, then $(c_g^\alpha)^2+\sum_{k=1}^\infty|c^\alpha_k|^2=\|e_\alpha\|_2^2$, $c_g^\alpha=\int_{\Omega_0}e_\alpha\phi_g=-\frac12\|e_\alpha\|_2^2$ and we can derive that
\begin{align*}
&\int_{\Omega_0} d
\vec{x} \bigg\{ \frac{1}{2}|\nabla e_\alpha|^2+(\beta+\beta_{12}) |\phi_g|^2|e_\alpha|^2+2\mu_g \phi_g e_\alpha
\bigg\}\\
&= \mu_g(c_g^\alpha)^2+\sum\limits_{k=1}^\infty\mu_k|c^\alpha_k|^2-\mu_g\|e_\alpha\|_2^2\\
&\ge (\mu_1-\mu_g)(\|e_\alpha\|_2^2-(c_g^\alpha)^2)\\
&=(\mu_1-\mu_g)\|e_\alpha\|_2^2(1-\|e_g^\alpha\|_2^2/4).
\end{align*}
Now, firstly, we need a lower bound for $\mu_1-\mu_g$, the so-called fundamental gap, which has been solved recently by
A. Ben and C. Julie \cite{fund}. Using equation (\ref{e-l}), applying elliptic theory with convex domain $\Omega_0$, it is easy to verify that $\phi_g\in H^2(\Omega_0)$ and hence belongs to $C^{0,\gamma}(\overline{\Omega_0})$ ($0<\gamma<\frac12$) by Sobolev embedding. Approximating $\Omega_0$ by convex domain $\Omega_\varepsilon$ (with smooth boundary) and applying Schauder estimates, we shall have $\phi_g\in C^{2,\gamma}(\overline{\Omega_\varepsilon})$ and  there exists some $c>0$ such that $|\phi_g|^2+c|\vec{x}|^2$ is convex (as Hessian matrix of $|\phi_g|^2$ is bounded by Schauder estimates). Hence, we can apply the results in Ref.~\cite{fund} to get  ($D_\varepsilon$ is the diameter of $\Omega_\varepsilon$)
\begin{equation}
\mu_1^\varepsilon-\mu_g^\varepsilon\ge \frac{3\pi^2}{D_\varepsilon^2},
\end{equation}
where $\mu_g^\varepsilon$ and $\mu_1^\varepsilon$ are the first and second eigenvalues, respectively, of $L_g$ in $H_0^1(\Omega_\varepsilon)$. By Min-max principles, letting $\varepsilon\to0$, we have  $\mu_g^\varepsilon\to\mu_g$ and $\mu_1^\varepsilon\to\mu_1$. Hence we find
\begin{equation}\label{diff}
\mu_1-\mu_g\ge \frac{3\pi^2}{D^2},
\end{equation}
where $D$ is the diameter of $\Omega_0$ [or if we assume $\Omega_0$ is a convex domain with smooth boundaries, (\ref{diff}) follows directly].

Secondly, we have  $\|e_\alpha\|_2^2\leq\int_{\Omega_0}d\vec{x}(|\phi_\alpha^g|^2+|\phi_g|^2)=2$.

Thirdly, we would like to derive $L^\infty$ bounds of $\phi_g$ and $\phi_\alpha^g$. The Euler-Lagrange equation for $\phi_\alpha^g$ reads as
\begin{equation}\label{e-l-2}
\mu_\alpha^g\phi_\alpha^g=-\frac12\nabla^2\phi_\alpha^g+\beta|\phi_\alpha^g|^2\phi_\alpha^g
+\beta_{12}|\phi_{\alpha^\prime}^g|^2\phi_\alpha^g,
\end{equation}
with $\alpha^\prime\neq\alpha $. For the nonlinear eigenvalues, we have the estimates $\mu_\alpha^g\leq 2E[\phi_1^g,\phi_2^g]\leq 2E_s[\phi_g]$, $\mu_g\leq E_s[\phi_g]$ and $E_s[\phi_g]$ can be bounded by choosing any test function (like the ground state of $-\Delta$), which gives $E_s[\phi_g]\leq \widetilde{C}(1+\beta)$ ($\widetilde{C}$ depends on $\Omega_0$).

If $\beta\ge1$, considering the point $x_0\in\Omega_0$ where $\phi_g$ attains its maximum, then $\Delta\phi_g(x_0)\leq0$ and from (\ref{e-l}), we have
\begin{equation*}
\mu_g\phi_g(x_0)\ge(\beta+\beta_{12})|\phi_g(x_0)|^2\phi_g(x_0),
\end{equation*}
which gives $\|\phi_g\|_{\infty}^2\leq \frac{\mu_g}{\beta+\beta_{12}}\leq2\widetilde{C}$. Similarly, we can obtain the $L^\infty$ bound for $\phi_\alpha^g$ using the Euler-Lagrange equation
 and $\|\phi_\alpha^g\|_{\infty}^2\leq \frac{\mu_\alpha^g}{\beta}\leq 4\widetilde{C}$. Thus, $\|\phi_g+\phi_\alpha^g\|_{\infty}^2\leq 12\widetilde{C}$. Combining the three observations above, we get
\begin{align*}
 E[\phi_1^g,\phi_2^g]-E[\phi_g,\phi_g]\ge
\sum\limits_{\alpha=1,2}\bigg\{\frac{3\pi^2}{2D^2}
-\frac{12\beta^\prime\widetilde{C}}{2} \bigg\} \|e_\alpha\|_2^2,
\end{align*}
which implies that for $0\leq\beta^\prime\leq\min\{\frac{\pi^2}{4D^2\widetilde{C}},1\}$, there must hold $e_\alpha=0$, i.e., $\eta=1$.

For $\beta\in[0,1]$, the approach above is not good. In this case, we see that $\mu_\alpha^g\leq 4\widetilde{C}$ and $\mu_g\leq 2\widetilde{C}$.
Using Sobolev inequality, in one dimension ($d=1$), we can find that
\begin{equation}\label{1d-C}
\|\phi_\alpha^g\|_{\infty}^2\leq\|\nabla\phi_\alpha^g\|_{2}\|\phi_\alpha^g\|_{2}\leq
\sqrt{\mu_\alpha^g}\leq 2\sqrt{\widetilde{C}}.
\end{equation}
Similarly, $\|\phi_g\|_{\infty}^2\leq \sqrt{2\widetilde{C}}$.
For two and three dimensions ($d=2,3$), recalling (\ref{e-l}) and (\ref{e-l-2}),
we can obtain from elliptic theory and Sobolev inequalities  that there exist  constants $C_1, C_2>0$ only depending on $\Omega_0$ such that $\|\phi_\alpha^g\|_{\infty}
\leq C_1\|\phi_\alpha^g\|_{H^2}\leq C_2\cdot\|\mu_\alpha^g\phi_\alpha^g-\beta|\phi_\alpha^g|^2\phi_\alpha^g
-\beta_{12}|\phi_{\alpha^\prime}^g|^2\phi_\alpha^g\|_{2}$, and $\|\phi_g\|_{\infty}\leq C_2
\|\mu_g\phi_g-(\beta+\beta_{12})|\phi_g|^2\phi_g\|_{2}$. In two and three dimensions, using Sobolev inequality, we have $\|\phi_\alpha^g\|_{6}\leq C_3\|\nabla\phi_\alpha^g\|_{2}\leq C_3\sqrt{\mu_\alpha^g}$ ($C_3$ depends on $\Omega_0$). Cauchy inequality leads to
\begin{equation*}
\|\phi_\alpha^g\|_{\infty}\leq C_2(\mu_\alpha^g+\beta\|\phi_\alpha^g\|_{6}^{3}
+\beta_{12}\|\phi_\alpha^g\|_{6}\|\phi_{\alpha^\prime}^g\|_{6}^2),
\end{equation*}
and thus $\|\phi_\alpha^g\|_{\infty}^2\leq C_4$ ($C_4$ depends on $\Omega_0$). Similarly, $\|\phi_g\|_{\infty}^2\leq C_5$ ($C_5$ depends on $\Omega_0$). Eventually, we have in all dimensions ($d=1,2,3$), there exists a constant $C_{\Omega_0}$ depending only on $\Omega_0$ such that
 $\|\phi_\alpha^g+\phi_g\|_\infty^2\leq C_{\Omega_0}$. Similar to the case with $\beta\ge1$, we have
\begin{align*}
& E[\phi_1^g,\phi_2^g]-E[\phi_g,\phi_g] \ge
\sum\limits_{\alpha=1,2}\left\{ \frac{3\pi^2}{2D^2}-
\frac{\beta^\prime C_{\Omega_0}}{2} \right\}\|e_\alpha\|_2^2,
\end{align*}
which leads to the conclusion that when $\beta^\prime<\min\{1,\frac{3\pi^2}{D^2C_{\Omega_0}}\}$, $\phi_\alpha^g=\phi_g$, i.e., $\eta=1$. In summary, for all $\beta\ge0$, if we choose
$\beta_{12}^c=\beta+\min\{1,\frac{3\pi^2}{D^2C_{\Omega_0}},\frac{\pi^2}{4D^2\widetilde{C}}\}>\beta$, then for all $0\leq\beta_{12}<\beta_{12}^c$, we shall have $\eta=1$.

\section{Phase separation as a spontaneous symmetry
breaking}\label{appSpon}

Consider a two-component BEC in a symmetric double-well
potential. Under the two-mode approximation, the mean-field
energy functional is
\begin{eqnarray}\label{double1}
  E &=& -J_a(\psi_{a1}^* \psi_{a2}+ \psi_{a2}^* \psi_{a1})-J_b(\psi_{b1}^* \psi_{b2}+ \psi_{b2}^* \psi_{b1})  \nonumber\\
  && +\frac{1}{2 }U_a (|\psi_{a1}|^4 + |\psi_{a2}|^4) + \frac{1}{2 }U_b(|\psi_{b1}|^4 + |\psi_{b2}|^4) \nonumber  \\
   && +V(|\psi_{a1}|^2|\psi_{b1}|^2 + |\psi_{a2}|^2|\psi_{b2}|^2).
\end{eqnarray}
Here $J_a$ and $J_b$ are the hopping amplitudes of the two
types of atoms, and $U_a$ and $U_b$ are the intra-component
onsite interaction strengths, while $V$ is the
inter-component one. The complex numbers $\psi_{a1}$ and
$\psi_{b1}$ ($\psi_{a2}$ and $\psi_{b2}$) are the
amplitudes of the two condensate wave functions on the left
(right) trap. They are constrained by the total atom
numbers, i.e., $|\psi_{a1}|^2 + |\psi_{a2}|^2 =N_a$ and
$|\psi_{b1}|^2 + |\psi_{b2}|^2 =N_b$. For the sake of
simplicity, in the following we shall assume $J_a=J_b=J\geq
0$, $U_a=U_b=U \geq 0$, and $N_a=N_b=N$. As far as the
ground state is concerned, it is legitimate to assume the
$\psi$'s real and positive. Therefore, we can write
$\psi_{a1}=\sqrt{N_{a1}}$, $\psi_{b1}=\sqrt{N_{b1}}$ and
similarly for other $\psi$'s.

First assume tunneling is turned off, i.e. $J=0$. Let
$N_{a1}= \frac{1}{2} N + \delta_a$ and $N_{b1}= \frac{1}{2}
N - \delta_b$. The energy (\ref{double1}) is
\begin{eqnarray}
  E(\delta_a, \delta_b) = U(\delta_a^2 +\delta_b^2)- 2V\delta_a \delta_b +
\text{const}.
\end{eqnarray}
It is readily determined that if $U>V$, the ground state is
of $\delta_a=\delta_b=0$. The two condensates are both
distributed evenly between the two wells, which is a
completely mixed configuration. If $U<V$ [the counterpart
of (\ref{cond}) in the present context], the ground state
is of $(\delta_a,\delta_b)=\pm (N/2,N/2)$, which
corresponds to complete phase separation---the two
condensates occupy the two wells separately. Therefore,
without tunneling, the miscibility-immiscibility transition
is a first-order phase transition with the critical point
being $V^c =U$.

Now turn on the tunneling. For the sake of simplicity,
suppose $\delta_a=\delta_b=\delta$. The energy as a
function of the order parameter $\delta$ is
\begin{eqnarray}
      E &=&-4J \sqrt{ \left(\frac{N}{2} \right)^2-
    \delta^2} + 2(U-V)\delta^2 +\text{const} \nonumber \\
   &=&  \left[\frac{4J}{N}+ 2(U-V) \right]\delta^2 + \frac{4J}{N^3}
\delta^4+   o(\delta^4)+ \text{const} .\quad\quad
\end{eqnarray}
Here we have the familiar Landau formalism for second order
phase transitions. The coefficient of the quartic term is
positive but the sign of the quadratic term changes from
positive to negative as $V$ surpasses the critical value
$V^c= U + 2J/N$. Corresponding, $\delta=0 $ is turned from
a minimum to a maximum point and phase separation develops.
Here we note that the tunneling, the kinetic term in the
present context, has two consequences. First, the
first-order transition is turned into a second-order one.
Second, the transition point is up shifted from $U$ to
$U+2J/N$. This is reasonable since phase separation costs
kinetic energy. What presented in Figs.~\ref{fig1} and
\ref{fig2} are parallel to these results but in continuum
(multi-mode) cases.

\section{Justification of the form of $\varphi_1$ in Eq.~(\ref{trial2})}
\label{appc}

In this Appendix, we show why among all the (degenerate) first excited states, the one in Eq.~(\ref{trial2}) is selected. For $d=2$, the ansatz more general than Eq.~(\ref{trial}) is
\begin{subequations}\label{trial3}
\begin{eqnarray}
    \phi_1 &=& c_0 \varphi_0 + c_x \varphi_x+ c_y\varphi_y,  \\
     \phi_2 &=& c_0
\varphi_0 - c_x \varphi_x- c_y \varphi_y,
\end{eqnarray}
\end{subequations}
with $\varphi_x= w_1(x)w_0(y)$, $\varphi_y= w_0(x)w_1(y)$, and $c_0$, $c_{x}$, $c_{y}$ being some real variables under the constraint $c_0^2 +c_x^2 +c_y^2 =1$. Substituting Eq.~(\ref{trial3}) into Eq.~(\ref{fun2}), we get the reduced energy functional $\widetilde{E}=E/(N\hbar^2/mL^2)$ as a function of $c_{x,y}$ as
\begin{eqnarray}\label{ecxcy}
\tilde{E}[c_x,c_y] &=& 2\pi^2 +\frac{9}{4}\beta_{12}+\left(3\pi^2 -\frac{15}{2}\beta_{12} \right)(c_x^2+c_y^2) \nonumber\\
& & +\frac{15}{2}\beta_{12}\left(c_x^2 +c_y^2 \right)^2+ \frac{3}{2}\beta_{12}c_x^2 c_y^2.
\end{eqnarray}
We see that for $\beta \leq \beta_{12}^c =\frac{2}{5}\pi^2$, the minimum is at $c_x=c_y=0$. For $\beta_{12} >\beta_{12}^c$, the minimum is no longer at the origin. However, for a fixed value of $c_x^2+c_y^2$, $\tilde{E}$ is minimized when the last term in Eq.~(\ref{ecxcy}) vanishes or when $c_x=0$ or $c_y=0$. That is why the particular ansatz in Eqs.~(\ref{trial}) and (\ref{trial2}) is appropriate and enough. We note that due to the symmetry of the trap, the reduced energy functional is invariant under the transform $(c_x,c_y) \rightarrow (\pm c_x,\pm c_y)$ and $(c_x,c_y) \rightarrow ( c_y, c_x)$. This symmetry is broken when phase separation occurs.

Similar analysis applies for $d=3$. In this case, the ansatz more general than Eq.~(\ref{trial}) is
\begin{subequations}\label{trial4}
\begin{eqnarray}
    \phi_1&=&  c_0 \varphi_0 + c_x \varphi_x+ c_y\varphi_y+c_z \varphi_z, \\
     \phi_2 &= & c_0\varphi_0 - c_x \varphi_x- c_y \varphi_y -c_z \varphi_z,
\end{eqnarray}
\end{subequations}
with $\varphi_x= w_1(x)w_0(y)w_0(z)$, $\varphi_y= w_0(x)w_1(y)w_0(z)$, $\varphi_z= w_0(x)w_0(y)w_1(z)$, and $c_0$, $c_{x}$, $c_{y}$, $c_z$ being some real variables under the constraint $c_0^2 +c_x^2 +c_y^2+c_z^2 =1$. Substituting Eq.~(\ref{trial4}) into Eq.~(\ref{fun2}), we get the reduced energy functional $\tilde{E} $ as a function of $c_{x,y,z}$ as
\begin{eqnarray}\label{ecxcycz}
\tilde{E} &=& 3\pi^2 +\frac{27}{8}\beta_{12}+\left(3\pi^2 -\frac{45}{4} \beta_{12}\right)(c_x^2+c_y^2+c_z^2)\quad\quad\ \nonumber\\
&&\quad\quad  +\frac{45 }{4} \beta_{12} \left(c_x^2 +c_y^2 +c_z^2\right)^2 \nonumber\\
&& \quad\quad  + \frac{9}{4} \beta_{12} \left(c_x^2 c_y^2 +c_y^2 c_z^2+c_z^2 c_x^2\right).
\end{eqnarray}
We see that for $\beta \leq \beta_{12}^c =\frac{4}{15}\pi^2$, the minimum is at $c_x=c_y=c_z=0$. For $\beta_{12} >\beta_{12}^c$, the minimum is no longer at the origin. However, for a fixed value of $c_x^2+c_y^2+c_z^2$, $\tilde{E}$ is minimized when the last term in Eq.~(\ref{ecxcycz}) vanishes or when two of the three $c$'s are zero. Again, we see that the particular ansatz in Eqs.~(\ref{trial}) and (\ref{trial2}) is appropriate and enough.


\begin{thebibliography}{99}
\bibitem{positive}
To avoid the possibility of collapse of the BEC under
attractive interaction, throughout this paper all the $g$'s
are assumed to be non-negative.
\bibitem{rabi}
Note that the scenario here is like that in \cite{malomed},
where the Rabi coupling between the two components favors
phase mixing and thus can suppress phase separation.
\end{thebibliography}
\end{document}